\makeatletter
\@ifundefined{@parse@version@dash}{%
\def\@parse@version#1{\@parse@version@0#1}
\def\@parse@version@#1/#2/#3#4#5\@nil{%
\@parse@version@dash#1-#2-#3#4\@nil}
\def\@parse@version@dash#1-#2-#3#4#5\@nil{%
  \if\relax#2\relax\else#1\fi#2#3#4 }
}{}
\makeatother

\documentclass[
aps,twocolumn,
groupedaddress,superscriptaddress,
amsmath,floatfix,nofootinbib
]{revtex4-2}

\usepackage{revsymb4-2}
\usepackage{graphicx}  
\usepackage{bm}

\begin{document}

\title{On channeling of charged particles in a single dielectric capillary}
  
\author{S.B. Dabagov}
\affiliation{INFN Laboratori Nazionali di Frascati, Via E. Fermi 54, I-00044 Frascati (RM), Italy}
  \affiliation{RAS P.N. Lebedev Physical Institute, Leninsky Pr. 53, 119991 Moscow, Russia}
  \affiliation{NR Nuclear University MEPhI, Kashirskoe Sh. 31, 115409 Moscow, Russia}
  
  \author{A.V. Dik}

  \affiliation{RAS P.N. Lebedev Physical Institute, Leninsky Pr. 53, 119991 Moscow, Russia}  

  \begin{abstract}
 
 We analytically analyse the motion of a nonrelativistic charged particle in a cylindrical single capillary. The effective potential for interaction of a charged particle with the inner surface of a capillary is derived as a sum of the averaged atomic potential of the capillary wall surface and the induced potential defined by collective surface excitations. We have shown that under certain conditions this potential becomes attractive and may hold a particle in a bound state due to  the surface excitations  that defines a so-called "surface channeling"  regime of motion. For the first time we have evaluated the induced potential revealing two limits to form either a well or a barrier that are delineated by the ratio of the insulator plasmon frequency to the frequency defined by the particle motion in a capillary.  
 
  \end{abstract}
  
 

\maketitle

\section{Introduction}

Controlling beams of charged particles, which determines the development in accelerator, storage-ring and collider techniques, had occupied the minds of physicists and engineers throughout the last century. Research in this area remains relevant to this day directing us to fundamental studies on basic knowledge of the origin and structure of the world (see the recent comprehensive review \cite{shizim-rmp2021}). The main principles of this branch of research are formulated through the interaction of charged particle beams in strong external electromagnetic fields, which can be created using solenoids, dipoles, quadrupoles, etc., magnets of various types, laser-plasma sources \cite{mizi-book2003,ziseho-nature2021}.

As known a strong electromagnetic interaction of charged particles takes also place in solids (dense media), as well as in proximity to their surfaces, sometimes offering the  field gradients far beyond our technical capabilities. The formation of strong fields of a certain geometry of small dimensions allows actually handling the particle beams, endowing them with requested characteristics for use in various experiments in both fundamental and applied research \cite{nato-asi1991}.

One of the promising methods for shaping the beams is the use of capillaries or capillary structures of various geometries. Despite the difference in the physics of the processes, a clear example is capillary x-ray optics (for the physics see in \cite{dabagov-physusp2003}), which has grown from a beautiful idea into a widely used x-ray radiation control tool. In fact, even before the development of capillary/polycapillary x-ray optics, in the early 1980s, the possibility of using curved reflecting surfaces as an alternative method for the formation of ion beams was discussed in a number of experimental works with theoretical estimates  based on the physics of charged particle channeling in crystals \cite{kumakhov-bookinru1986}. The interaction of an incident ion with a reflecting surface was described in the approximation of a continuous surface atomic potential, confirming very small angles of ion reflection from a smooth surface regardless of the change in the beam charge characteristics. In order to increase the angles of deflection of the beams, to use curved surfaces, in particular, dielectric capillaries, was proposed (\cite{kuko-physrep1990} and Refs. therein). However, these studies were not properly continued, while the further use of capillaries (capillary bundles) is presently known in relation to the beams of x rays and thermal neutrons \cite{khomac-xoi2010,dagl-rpc2019}.

On the contrary, since the first observations of the ions passage through dielectric capillaries \cite{stol}, this topic has attracted a growing attention of physicists. The phenomenon is explained by the fact that when the ion beam is transmitted by the capillary, at the beginning, the part of the beam settles on the inner capillary surface and, having accumulated there, creates a repulsive potential, which becomes a channel-guiding potential for a charged particle \cite{stol, schi}. 

The undoubted interest in the development of new methods for handling particle beams based on capillary systems has resulted in a wide spectrum of experimental activities dedicated to this phenomenon, while theoretical works counts disproportionally much less published articles (see in the detailed review \cite{sto-ya_phrep2016} and  Refs. therein). Theoretical studies have mainly aimed at simulating the motion of charged particle beams inside capillary insulators using the Monte Carlo method \cite{schi, schi1}.

In this work, the interaction of a charged particle with the inner wall of a dielectric capillary has been analytically evaluated, assuming the wall thickness to be large for creating collective features in the interaction potential. In addition, our approach shortens and slightly simplifies general known calculations without losing the subtle features of the particle interaction with the inner capillary wall.

\section{Surface excitation}

\begin{figure}
\begin{center}
\includegraphics[width=5cm,draft=false]{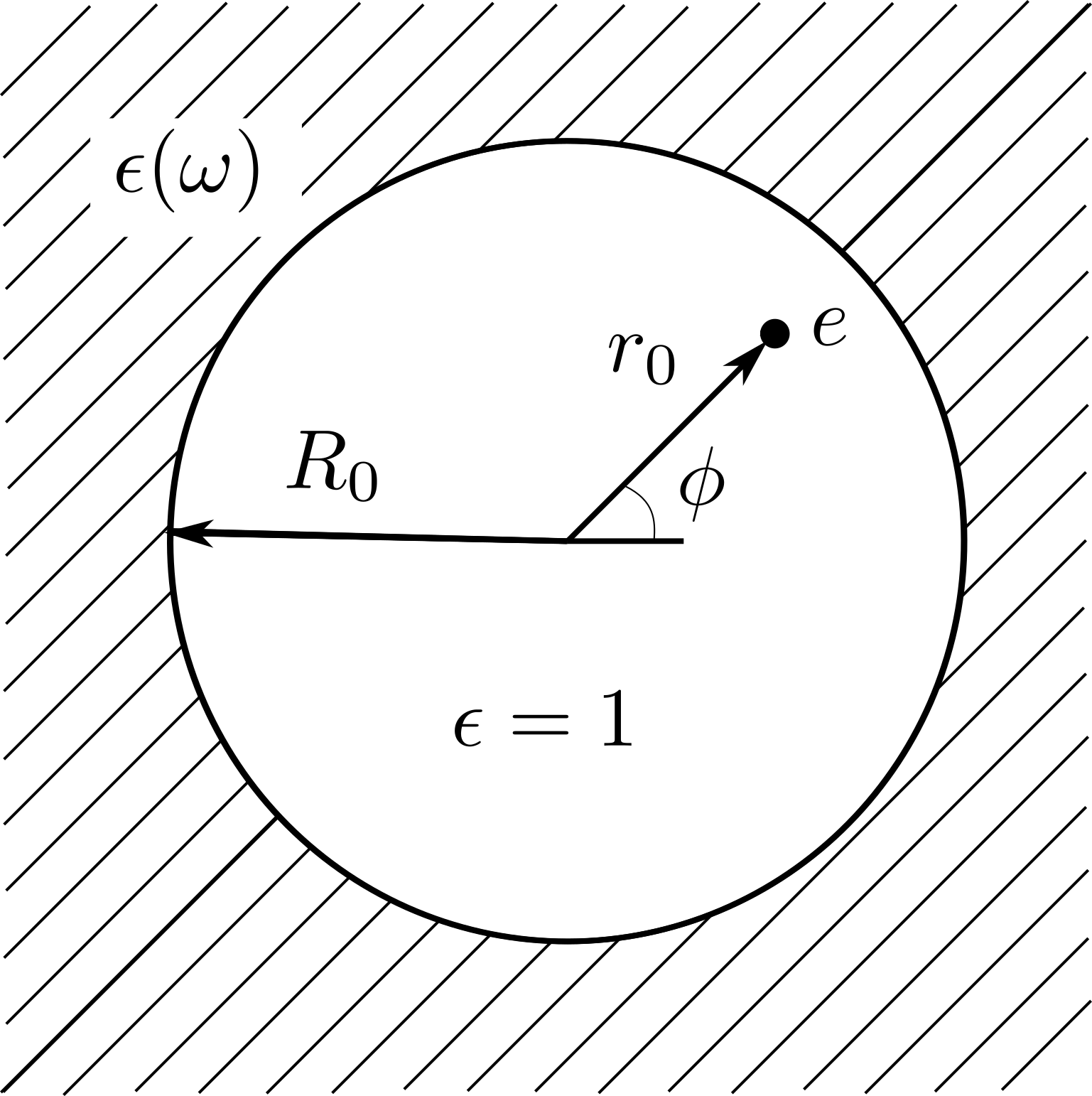}
\end{center}
\caption{Scheme of a cylindric cavity of radius $R_0$ in the infinite dielectric, characterised by the permittivity $\epsilon(\omega)$, with a particle of the charge $e$ moving inside the cavity. For the cavity we choose $\epsilon = 1$.}\label{fig1.0}
\end{figure}

Let us consider the electromagnetic field in a nonmagnetic substance of the permittivity $\epsilon(\omega)$ that occupies the infinite space with a cylindrical cavity of radius $R_0$ inside (Fig.~\ref{fig1.0}). Then, in absence of free currents and charges, based on Maxwell's equations, the Fourier components of the field scalar and vector potentials are defined by the following equation
\begin{equation}
\label{eq1}\left(\nabla^2_{\bot}+\frac{\partial^2}{\partial z^2}+\frac{\omega^2}{c^2}\epsilon(\omega)\right)\left(\begin{array}{ll}\varphi_{\omega}\\ \mathbf{A}_{\omega}\end{array}\right)=0\:,
\end{equation}
where $\nabla^2_{\bot}$ is the part of the Laplace operator of the coordinates transverse to the $Oz$ axis, which coincides with the cavity axis. The field potentials satisfy the following gauge relation
\begin{equation}
\label{eq2}\nabla\mathbf{A}_{\omega}-i\frac{\omega}{c}\epsilon(\omega)\varphi_{\omega}=0
\end{equation}
As we are interested in characterising the damping surface waves excited at the boundary cavity interface, the solution of Eq.(\ref{eq1}) under the condition (\ref{eq2}) in a cylindrical coordinate system can be presented in the form (for ease of perception the index $\omega$ is omitted)
\begin{equation}
\label{eq3}\varphi=\sum\limits_{q,n}\varphi^0_nf_n(\nu r)e^{i(qz+n\phi)}\: ,
\end{equation}
\begin{equation}
\label{eq3.1}A_z=\sum\limits_{q,n}A^0_nf_n(\nu r)e^{i(qz+n\phi)}\: ,
\end{equation}
\[A_{r}=\sum\limits_{q,n}\left[-i\frac{q}{\nu}\left(qA_n^0+\frac{\omega}{c}\epsilon(\omega)\varphi_n^0\right)f_n'(\nu r)+\right.\]
\begin{equation}
\nonumber
\label{eq4}\left.+\frac{1}{r}B_n^0f_n(\nu r)\right]e^{i(qz+n\phi)}\: ,
\end{equation}
\[A_{\phi}=\sum\limits_{q,n}\left[\frac{n}{\nu^2r}\left(qA_n^0+\frac{\omega}{c}\epsilon(\omega)\varphi_n^0\right)f_n(\nu r)+\right.\]
\begin{equation}
\nonumber
\label{eq5}\left.+i\frac{\nu}{n}B_n^0f'_n(\nu r)\right]e^{i(qz+n\phi)}\: ,
\end{equation}
\begin{equation}
\label{eq6}f_n(\nu r)=\left\{\begin{array}{ll}K_n(\nu r),\nu=\sqrt{q^2-\epsilon \omega^2/c^2}, r>R_0 \\ I_n(\nu r),\nu=\sqrt{q^2-\omega^2/c^2}, r\le R_0 \end{array}\right.
\end{equation}
Here $\epsilon(\omega)=1$ inside the hollow cavity $r<R_0$, $\{\varphi^0_n, A_n^ 0, B_n^0\}$ are the constants, $n$ is the integer, $K_n(\xi),I_n(\xi)$ are the modified Bessel functions. From the condition of continuity of the components of the electric and magnetic fields tangential to the interface, we derive the dispersion law $q(\omega)$ in the new notation $x=qc/\omega$ and $\alpha=R_0\omega/c$ 
\[\frac{x^2n^2(\epsilon-1)^2}{\alpha^2(x^2-1)(x^2-\epsilon)}=\left[\sqrt{x^2-1}\frac{K'_n(\alpha\sqrt{x^2-\epsilon})}{K_n(\alpha\sqrt{x^2-\epsilon})}-\right.\]
\[\left.-\sqrt{x^2-\epsilon}\frac{I'_n(\alpha\sqrt{x^2-1})}{I_n(\alpha\sqrt{x^2-1})}\right]\left[\epsilon\sqrt{x^2-1}\frac{K'_n(\alpha\sqrt{x^2-\epsilon})}{K_n(\alpha\sqrt{x^2-\epsilon})}-\right.\]
\begin{equation}
\label{eq7}\left.-\sqrt{x^2-\epsilon}\frac{I'_n(\alpha\sqrt{x^2-1})}{I_n(\alpha\sqrt{x^2-1})}\right]
\end{equation} 

\begin{figure}[h]
\begin{center}
\includegraphics[width=6.5cm,draft=false]{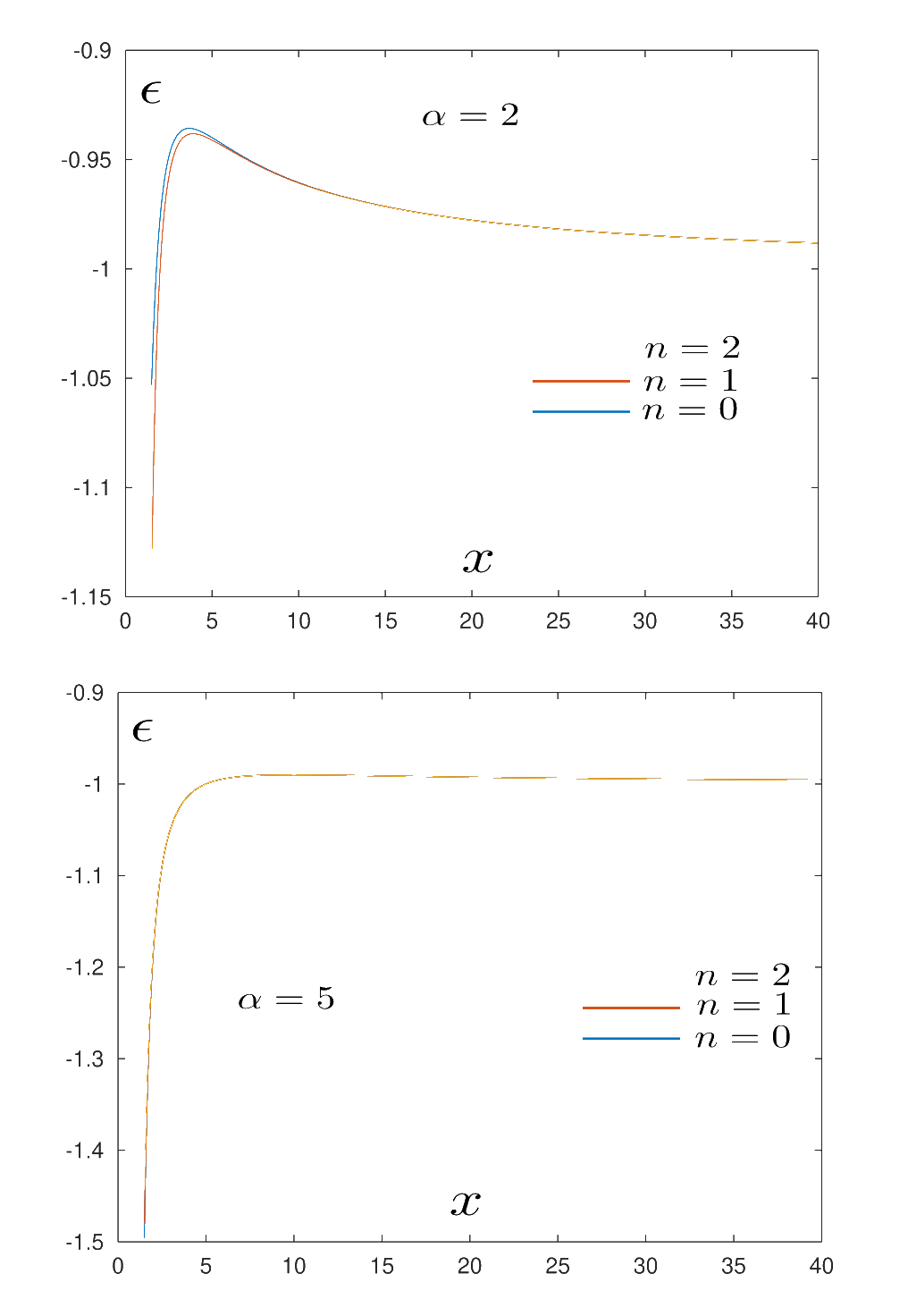}
\end{center}
\caption{The solutions of the disperse equation (\ref{eq7}) for two values of the dimensionless parameter $\alpha$, which characterises the ratio of the cavity radius to the wavelength of surface excitations, and the first three orders of Bessel function. Both dependences reveal identical evolution with an increase in the values $\alpha$ and $n$, while the valid values of  $\epsilon(\omega)$ are essentially different. \label{num}}
\end{figure}

Solving analytically Eq.(\ref{eq7}) is a rather routine and complex procedure. However, we can solve it numerically (Fig.~\ref{num}). The curves of the figure present two various solutions for different values of the parameter $\alpha$ and the first orders of Bessel function. As seen, the equation (\ref{eq7}) can be solved only at negative $\epsilon(\omega)$ in very narrow interval near the point $\epsilon(\omega) = -1$. At higher values of the parameter $\alpha$ and/or the Bessel function order we get the dependences presented in the bottom plot of Fig.~\ref{num} but with the difference that the range of valid values $\epsilon(\omega)$ rapidly decreases and tends to the value $\epsilon(\omega) = -1$.
Thus, neglecting the wavelengths of surface excitations of the order of the cavity radius $\lambda\sim R_0$, we can conclude that the wave-vector of surface excitations should be within the range $0\le q\le q_0$ ($q_0$ is the wave-vector corresponding to cut-off frequency), while the corresponding frequency $\omega_s$ can be reduced from the solution of the equation $\epsilon(\omega_s) = -1$.

Let examine now two physically significant limiting cases determined by the parameter $\alpha = R_0/\lambda_s$, the ratio of the resonator radius $R_0$ to the wavelength of surface excitations $\lambda_s$. For $\alpha\ll 1$, Eq.(\ref{eq7}) can be solved only at $\epsilon(\omega)=-1$, while for $\alpha\gg1$ the solution can be presented in the form of a series expansion in powers of a small parameter $(1/\alpha)$ 

\begin{equation}
\label{eq8}x=\sqrt{\frac{\epsilon}{1+\epsilon}}+\frac{1}{2\alpha}\frac{\epsilon-1}{\sqrt{\epsilon(1+\epsilon)}}+...
\end{equation} 
For large $\alpha$, i.e. for the wavelengths $\lambda_s\ll R_0 $, the dispersion relation $q (\omega) $ slightly differs from the case of a flat reflecting plane \cite{amer}. Thus, if we neglect the wavelengths $\lambda_s=c/ \omega_s\sim R_0 $, the frequency of surface excitations $\omega_s $ is assumed to be constant, while the wave vector varies within the range $ 0 \le q \le q_0 $.

At these assumptions, the charged particle interaction with the cavity surface is described by the scalar potential operator of surface excitations 
\[\hat{\Phi}(\mathbf{r})=\hbar\sum\limits_{q,s}f_{s}(qr)\times\]
\begin{equation}
\label{eq9}\times\left(g_{qs}\hat{c}_{qs}e^{i(qz+s\phi)}+g^*_{qs}\hat{c}^+_{qs}e^{-i(qz+s\phi)}\right)\:,
\end{equation}
where the excitation amplitudes $ g_{qs} $ are calculated in Appendix \ref{app1}, while the function $ f_s(qr) $ is defined by Eq.(\ref{eq6}) at $ \nu\approx q $.

\section{Interaction potential}

Since we deal with a nonrelativistic charge moving in a cylindrical cavity, the potential energy of interaction with the atomic system of a dielectric can be calculated as the sum of Coulomb potentials, represented as a Fourier series, split at the cut-off frequency $ q_0 $ \cite{madelung}
\[U_{int}(\mathbf{r})=\frac{4\pi}{V_g}\sum\limits_{\mathbf{r}_j}\left(\sum\limits_{\mathbf{q}>\mathbf{q}_0}\frac{e_pe_j}{q^2}e^{i\mathbf{q}(\mathbf{r}_p-\mathbf{r}_j)}+\right.\]
\begin{equation}
\label{eq1.0.1}\left.+\sum\limits_{\mathbf{q}<\mathbf{q}_0}\frac{e_pe_j}{q^2}e^{i\mathbf{q}(\mathbf{r}_p-\mathbf{r}_j)}\right)\: ,
\end{equation}
where $e_{p(j)},\mathbf{r}_{p(j)}$ are, respectively, the charges and radius-vectors of the particle $p$ and the surface nuclei and electrons $j$, $V_g$ is the normalisation volume. The 1st term describes the interaction at small distances  $l<q_0^{-1}$ that does not allow the averaging at calculations \footnote{For instance, for metals,  $q_0^{-1}$ is of the order of the atom size \cite{ash}.} , while the 2nd - at $l>q_0^{-1}$, i.e. the interaction with collective surface excitations, can be evaluated by averaging single interaction features over all the system. As aforementioned, the latter is described by the potential operator (\ref{eq9}). The 1st term of Eq.(\ref{eq1.0.1}), in turn, is described as the sum of potential energies $V(\mathbf{r}_p)$ for the 'particle - screened atomic' interaction over all surface atoms \cite{ded} that in view of the equation of motion for a particle transforms into the averaged atomic potential of the surface.

Generally, for a particle of charge $e$ and mass $m$ moving inside a cylindrical hollow cavity formed by an infinite insulator we can write the Hamiltonian in the form 
\[\hat{H}=\int\hat{\psi}^{+}(\mathbf{r},z)\left(-\frac{\hbar^2}{2m}\nabla^2+V(\mathbf{r},z)\right)\hat{\psi}(\mathbf{r},z)d^2rdz+\]
\begin{equation}
\label{eq1.1}+\hat{H}_s+\int \hat{\psi}^{+}(\mathbf{r},z)\hat{\Phi}(\mathbf{r},z)\hat{\psi}(\mathbf{r},z)d^2rdz\:,
\end{equation}
where $\mathbf{r}=(r,\phi)$ is the radius-vector in the transverse plane $xOy$, while the $Oz$-axis coincides with the cavity longitudinal axis, $\hat{\psi}$ is the particle field operator (fermion-operator), $\hat{\Phi}$ is the operator describing the particle interaction with the surface excitations (boson-operator). The potential energy of the atom-particle interaction $V(\mathbf{r})$ corresponds to the 1st term of Eq.(\ref{eq1.0.1}), while the operator $\hat{H}_s$ - to the energy of non-interacting surface excitations. Here, the particle field operator $\hat{\psi}$ is defined by the sum over the quantum states of the motion
\begin{equation}
\label{eq1.3}\hat{\psi}(\mathbf{r},z)=\sum\limits_{k,n,l}\varphi_{knl}(\mathbf{r},z)\hat{a}_{knl}\: ,
\end{equation}
where $\hat{a}_{knl}$ is the annihilation operator, $k$ is the z-projection of the charged particle wave vector, $n,\:l$ are the radial and azimuthal quantum numbers ($\mathbf{n}=(n,l)$). The wave function $\varphi_{knl}$ can be divided into longitudinal and transverse components
\begin{equation}
\label{eq1.4}\varphi_{knl}(\mathbf{r},z)=\frac{1}{\sqrt{L}}u_{nl}(\mathbf{r})e^{ikz}
\end{equation}
with the cavity longitudinal size $L$. At these definitions the interaction operator can be reduced to a simpler operator equation 
\[\hat{H}_i=\hbar\sum\limits_{q,s,k,\mathbf{n},\mathbf{n}'}\left(\Gamma_{\mathbf{n}'\mathbf{n}}(q,s)\hat{c}_{qs}\hat{a}^+_{k+q,\mathbf{n}'}\hat{a}_{k\mathbf{n}}+\right.\]
\begin{equation}
\label{eq1.5}\left.+\overline{\Gamma}_{\mathbf{n}'\mathbf{n}}(q,s)\hat{c}^+_{qs}\hat{a}^+_{k\mathbf{n}}\hat{a}_{k+q,\mathbf{n}'}\right)\:,
\end{equation}
where
\begin{equation}
\label{eq1.6}\Gamma_{\mathbf{n}'\mathbf{n}}(q,s)=g_{qs}\int f_{qs}(r)e^{is\phi}u^*_{\mathbf{n}'}(\mathbf{r})u_{\mathbf{n}}(\mathbf{r})d^2r
\end{equation}
The time evolution of the boson operators $\hat{c}_{qs}$ in Eq.(\ref{eq1.5}) can be estimated introducing new operators  $\hat{C}_{qs}=\hat{c}_{qs}e^{i\omega_{qs}t}$, $\hat{C}^+_{qs}=\hat{c}^+_{qs}e^{-i\omega_{qs}t}$ and $\hat{A}_{k\mathbf{n}}=\hat{a}_{k\mathbf{n}}e^{i\Omega_{\mathbf{n}}(k)t}$, $\hat{A}^+_{k\mathbf{n}}=\hat{a}^+_{k\mathbf{n}}e^{-i\Omega_{\mathbf{n}}(k)t}$ for the quantum energies of both surface excitations $\hbar\omega_{qs}$ and particle $\hbar\Omega_{\mathbf{n}}(k)$. For getting the expression of effective Hamiltonian $\hat{H}_{eff}$ we have followed the discussion of known work \cite{jap}, where the calculation procedure was in detail described. The only difference is in the particle freedom, i.e. in our case the particle is free in one direction, while in \cite{jap} - in two ones. Hence, in further we can briefly point out just the main steps.

Assuming the weak interaction we deal with, the new boson operators should vary in time much faster than those of fermion. That allows easy integrating the Heisenberg equation that results in the following expressions for the boson operators

\[\hat{C}_{qs}(t)=\sum\limits_{k,\mathbf{n},\mathbf{n}}\overline{\Gamma}_{\mathbf{n}'\mathbf{n}}(q,s)\hat{A}^+_{k\mathbf{n}}\hat{A}_{k+q,\mathbf{n}'}\times\]
\begin{equation}
\label{eq1.7.11}\times\frac{e^{-i(\Omega_{\mathbf{n}'}(k+q)-\Omega_{\mathbf{n}}(k)-\omega_{q,s})t}}{\Omega_{\mathbf{n}'}(k+q)-\Omega_{\mathbf{n}}(k)-\omega_{q,s}-i\delta}\:,
\end{equation}
\[\hat{C}^+_{qs}(t)=\sum\limits_{k,\mathbf{n},\mathbf{n}}\Gamma_{\mathbf{n}'\mathbf{n}}(q,s)\hat{A}^+_{k+q,\mathbf{n}'}\hat{A}_{k,\mathbf{n}}\times\]  
\begin{equation}
\label{eq1.8.11}\times\frac{e^{i(\Omega_{\mathbf{n}'}(k+q)-\Omega_{\mathbf{n}}(k)-\omega_{q,s})t}}{\Omega_{\mathbf{n}'}(k+q)-\Omega_{\mathbf{n}}(k)-\omega_{q,s}+i\delta}
\end{equation}  
Using these expressions the effective Hamiltonian for the interaction of a charged particle with the field of surface excitations can be presented as follows
\begin{equation}
\label{eq1.7}\hat{H}_{eff}=\sum\limits_{k,q,\mathbf{n},\mathbf{n}',s,\mathbf{j}}\frac{\hbar \Gamma_{\mathbf{n}',\mathbf{n}}(q,s)\overline{\Gamma}_{\mathbf{j},\mathbf{n}}(q,s)\hat{a}^+_{k\mathbf{n}'}\hat{a}^{}_{k\mathbf{j}}}{\Omega_{\mathbf{n}'}(k)-\Omega_{\mathbf{n}}(k-q)-\omega_{qs}+i\delta}
\end{equation}
 In our case the particle longitudinal energy is much greater than the transverse one $E_{\parallel}=\hbar^2k^2/(2m)\gg \hbar\omega_{\bot}^{n,l}$. Hence, the denominator in Eq.(\ref{eq1.7}) becomes equal to $\triangle(k,q,s)=(\hbar/m)kq-(\hbar/2m)q^2-\omega_{q,s}+i\delta$, and, since $\sum_{\mathbf{n}}\left<u_{\mathbf{n}}(\mathbf{r}')\left|\right.u_{\mathbf{n}}(\mathbf{r})\right>=\delta(\mathbf{r}-\mathbf{r}')$, the effective Hamiltonian is defined by the matrix element
\begin{equation}
\label{eq1.8}\hat{H}_{eff}=\int \hat{\psi}^+(\mathbf{r})\Phi(k,r)\hat{\psi}(\mathbf{r})d^2rdz\:,
\end{equation}
where 
\begin{equation}
\label{eq1.8.1}\Phi(k,r)=\sum\limits_{q,s}\frac{\hbar\left|g_{qs}f_{qs}(r)\right|^2}{\triangle(k,q,s)}
\end{equation}
The real part of Eq.(\ref{eq1.8}) is responsible for elastic processes in a complex potential energy, while its imaginary part - for  inelastic ones. 

The change in the energy of surface excitations due to the interaction with the particle can be estimated calculating the time derivative of the operator of the energy of surface excitations
\begin{equation}
\label{eqadd2.1}\frac{\partial \hat{E}_s}{\partial t} = \sum\limits_{q,s} \hbar\omega_{qs}\frac{\partial}{\partial t} \hat{c}^+_{qs}(t)\hat{c}_{qs}(t)
\end{equation} 
For that we use the expressions (\ref{eq1.7.11}), (\ref{eq1.8.11}), which describe the time evolution for the boson operators. Substituting them into (\ref{eqadd2.1}) and neglecting the recoil term $\frac{\hbar}{2m}q^2$, we obtain
\[\frac{\partial \hat{E}_s}{\partial t} = \sum\limits_{q,s,\mathbf{n},\mathbf{n}',\mathbf{n}_1,\mathbf{n}_1',k,\kappa} i\hbar\omega_{qs}\overline{\Gamma}_{\mathbf{n}'\mathbf{n}}\Gamma_{\mathbf{n}_1'\mathbf{n}_1}\hat{A}^{+}_{k,\mathbf{n}}\hat{A}_{k+q,\mathbf{n}'}\times\]
\[\times \hat{A}^{+}_{\kappa + q,\mathbf{n}_1'}\hat{A}_{\kappa,\mathbf{n}_1}\left(\frac{1}{\frac{\hbar}{m}kq - \omega_{qs}-i\delta} - \right.\]
\begin{equation}
\label{eqadd2.2}\left.-\frac{1}{\frac{\hbar}{m}\kappa q - \omega_{qs}+i\delta}\right)e^{i\frac{\hbar}{m}(\kappa - k)q t}
\end{equation} 
In this expression we assumed the time dependence for the fermion operators to be mostly defined by the exponent term. If now we apply the following rule 
\begin{equation}
\label{eqadd2.3}\frac{1}{x - a \pm i\delta} = \mathcal{P}\frac{1}{x-a} \mp i\pi\delta(x-a)
\end{equation}
and take into account that at large times only the terms with $\kappa = k$ make a notable contribution, then the expression (\ref{eqadd2.2}) can be reduced to the following
\[\frac{\partial \hat{E}_s}{\partial t} = -2\pi\sum\limits_{q,s,\mathbf{n},\mathbf{n}',\mathbf{n}_1,\mathbf{n}_1',k} \hbar\omega_{qs}\overline{\Gamma}_{\mathbf{n}'\mathbf{n}}\Gamma_{\mathbf{n}_1'\mathbf{n}_1}\times\]
\begin{equation}
\label{eqadd2.4}\times \hat{A}^{+}_{k,\mathbf{n}}\hat{A}_{k+q,\mathbf{n}'}\hat{A}^{+}_{k + q,\mathbf{n}_1'}\hat{A}_{k,\mathbf{n}_1}\delta\left(\frac{\hbar}{m}kq - \omega_{qs}\right)
\end{equation}
Assuming that at the initial moment the particle was in a state with quantum numbers $ (k, \mathbf{n}) $, the particle energy loss can be defined considering the time derivative of the particle energy to be equal to that of surface excitations with the opposite sign (hereinafter, the longitudinal velocity $ v = \frac{\hbar}{m}k $ is introduced)
\begin{equation}
\label{eqadd2.5}\frac{\partial E_p}{\partial t} = -2\pi\sum\limits_{q,s} \hbar\omega_{qs}\left|\Gamma_{\mathbf{n}\mathbf{n}}(q,s)\right|^2\delta\left(vq - \omega_{qs}\right)
\end{equation}
To obtain an expression for the particle energy loss as a function of its coordinate, let assume that $\left|u_{\mathbf{n}}^*u_{\mathbf{n}}\right|^2 = \delta(r - r_0)\delta(\phi - \phi_0)/r$, i.e. the particle is at the point $ (r_0, \phi_0) $. Moreover, taking into account the dispersion expression for surface excitations, namely, the constancy of the frequency $ \omega_s $ for any $ q<q_0 $ and passing from the sum over $ q $ to the integral, we obtain the following expression for the energy loss of a particle in dependence on its position (to avoid confusion, the index $ s $ in (\ref{eqadd2.5}) is substituted by $ n $)
\begin{equation}
\label{eqadd2.6}\frac{\partial E_p}{\partial t} = -\frac{L}{v}\hbar\omega_{s}\sum\limits_n\left|g_{qn}f_{qn}(r_0)\right|^2_{q = \omega_s/v}
\end{equation}
For instance, the particle energy loss due to the interaction with surface plasmons equals to
\begin{equation}
\label{eq1.8.2}-\frac{dE_p}{dz}=\left(\frac{e\omega_s}{v}\right)^2\sum\limits_{n=-\infty}^{\infty}\frac{K_n\left(\frac{\omega_s}{v}R_0\right)}{I_n\left(\frac{\omega_s}{v}R_0\right)}I_n^2\left(\frac{\omega_s}{v}r_0\right)\,,
\end{equation}
In the limiting case $R_0\to\infty$, Eq.(\ref{eq1.8.2}) defines the energy loss of a charged particle moving along the plane \cite{amer2} (see Appendix \ref{app2})
\begin{equation}
\label{eq1.8.3}-\frac{dE}{dz}=e^2\left(\frac{\omega_s}{v}\right)^2K_0\left(2\frac{\omega_s}{v}x\right)\:,
\end{equation}
where $x=R_0-r_0$ is the particle-to-plane distance.

Ones we use Eq.(\ref{eq1.8}), the equation of transverse motion inside a cavity can be written in the following way

\[\left(-\frac{\hbar^2}{2m}\nabla^2_{\bot}+\frac{1}{L}\int V(\mathbf{r}_{\bot},z)dz+\Phi(k,r)\right)u_{\mathbf{n}}(\mathbf{r}_{\bot})=\]
\begin{equation}
\label{eq1.9}=\hbar\omega_{\bot}^{\mathbf{n}}u_{\mathbf{n}}(\mathbf{r}_{\bot})\,,
\end{equation}
which is actually describes the channeled motion. Indeed, neglecting the change in the particle longitudinal momentum results in averaging the atomic potential over the surface plane. The surface potential is calculated by summing the potentials of individual atoms, where the number of atoms per unit length along the cavity axis is much greater than unity $N=n2\pi R_0\gg1$. After integration over the surface, the 2nd term in Eq.(\ref{eq1.9}) becomes dependent only on the distance from the cavity axis $ (1/L)\int V(\mathbf{r}_{\bot},z)dz \equiv \bar{V}(r)$.

Solving the equation for elastic processes, we consider only the real part of induced potential, while the potential imaginary part leads to a finite lifetime for each quantum state. Thus, the induced potential with the amplitude of surface excitations $g_{qn}=g_n(qR_0)$ is defined by

\[U_{ind}(r)\equiv Re\left[\Phi(v,r)\right]=\]
\begin{equation}
\label{eq1.11}=\frac{L}{2\pi}\sum\limits_{n=-\infty}^{\infty} \mathcal{P} \int\limits_0^{q_0}\frac{\hbar \left|g_n\left(qR_0\right)\right|^2}{vq-\omega_s}I_n^2\left(qr\right)dq\:,
\end{equation}
where $\mathcal{P}$ means the principal value of the integral. 

Hence, we obtain the effective potential for a channeled particle inside a cylindrical cavity as a sum of averaged and induced potentials $U_{eff}(r)=\bar{V}(r)+U_{ind}(r)$. The 1st term is evaluated as a continuous potential via known technique in channeling physics \cite{lindhard1965,gemmell-rmp1974} and does not result in any unexpected behaviours, while the 2nd one brings us to new features.

An approximate analysis of general induced potential for two extreme cases, i.e. $ \omega_sR_0/v\gg1$ and $\omega_sR_0/v\ll1$, is shown in Fig.~\ref{fig2}. At $ \omega_sR_0/v\ll1$, the induced potential is defined as follows (see Appendix \ref{app3}) \footnote{All the results stated below are valid under the condition $ q_0v> \omega_s $, which corresponds to the presence at least one singularity within the integration interval in Eq.(\ref{eq1.11}).}
\begin{figure}[h]
\begin{center}
\includegraphics[width=6.5cm,draft=false]{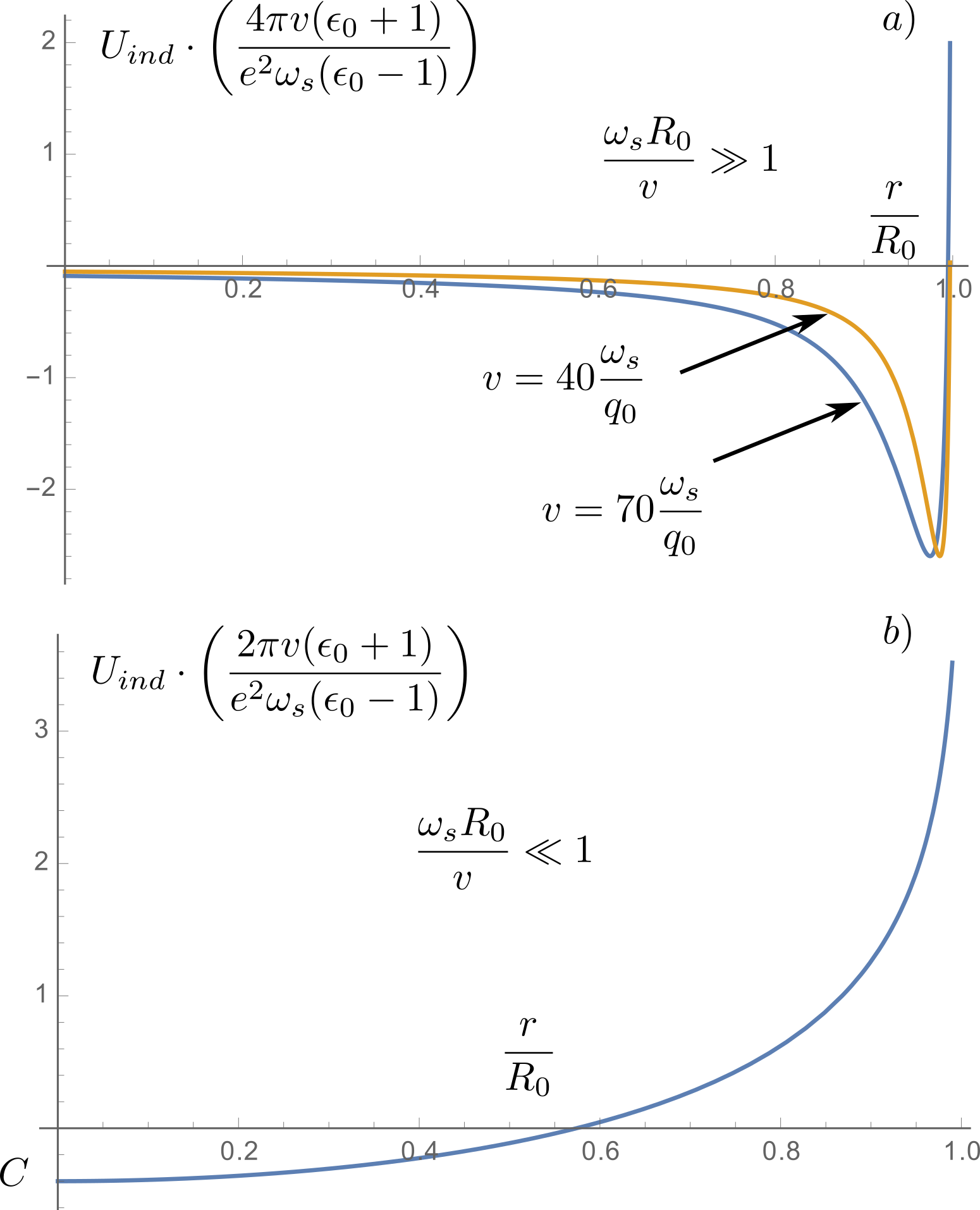}
\end{center}
\caption{Dimensionless induced potential versus the distance from the cavity axis. a). The graph at the condition $\omega _s R_0 / v \gg 1$ shows two curves corresponding different values of longitudinal velocity. As seen, at the particle velocity increase the potential shape does not change, and the curve minimum tends to the cavity center. b). The general induced potential at $\omega _s R_0 / v \ll 1$, while a concrete plotting is for $q_0 R_0 = 10^3$ and $\omega _s = 10^{15}$~c$^{-1}$ with the characteristic value $q_0 \sim 10^7$~c$^{-1}$ corresponding the cavity radius $R_0 \sim 10^{-4}$~cm.}\label{fig2}
\end{figure}
\[U_{ind}(r)= -\frac{e^2\omega_s}{2 \pi v}\frac{\epsilon_0-1}{(\epsilon_0+1)}\times\]
\begin{equation}
\label{eq1.12.1.1}\times\ln{\left(\frac{q_0v}{\omega_s}-1\right)}\ln{\left(1-\frac{r^2}{R_0^2}\right)} + C
\end{equation}
with
\[C=\frac{\epsilon_0-1}{\epsilon_0}\left[\ln{\left(\frac{q_0v}{\omega_s}-1\right)}\left(\ln{\left(\frac{\omega_sR_0}{2v}\right)}+\gamma_0\right)-\right.\]
\begin{equation}
\label{eq1.12.1.2}\left.-\left(\mathrm{Li}_2\left(1-\frac{q_0v}{\omega_s}\right)-\mathrm{Li}_2\left(1\right)\right)\right]\:,
\end{equation}
where $ \gamma_0 $ is the Euler's constant, $ \mathrm {Li} _2 (\xi) $ is the polylogarithm function. 
On the contrary, at $ \omega_sR_0 / v \gg 1 $ the induced potential is evaluated by the following expression  (see Appendix \ref{app3}) \footnote{Here, for clarity of the result, we set $ q_0v \gg \omega_s $.}

\begin{equation}
\label{eq1.12}U_{ind}(r)= -\frac{e^2\omega_s}{4 \pi v}\frac{\epsilon_0-1}{(\epsilon_0+1)}G\left(\frac{3\omega_sR_0}{2v}\left(1-\frac{r}{R_0}\right)\right)
\end{equation}
where the function
\begin{equation}
\label{eq1.12.1}G\left(\xi\right)=\int\limits_{0}^{\infty}\mathrm{Ei}\left(\xi\left(1+\frac{\mathrm{ch}{(t)}}{3}\right)\right)e^{-\xi(1+\mathrm{ch}{(t)}/3)}dt\:,
\end{equation}
 reaches its maximum $G(\xi_m)=2.59815$ at $\xi_m=0.545901$ ($\mathrm{Ei}(\xi)$ is the exponential integral function). That allows obtaining the potential minimum position  
\begin{equation}
\label{eq1.12.3}\left.\frac{r}{R_0}\right|_{max}= 1-0,3639\cdot\frac{v}{\omega_sR_0}
\end{equation}
with the minimum value of the induced potential
\begin{equation}
\label{eq1.12.3.1.1}U^{min}_{ind} = -0.2\cdot\frac{e^2\omega_s}{v}\frac{\epsilon_0-1}{(\epsilon_0+1)}
\end{equation}
If the particle velocity increases, the potential minimum position very slowly shifts closer to the cavity center, keeping the potential shape itself  practically unchanged. At the cavity surface the induced potential logarithmically diverges $U_{ind}(r\to R_0)\sim\ln{(R_0-r)}$, while at the cavity center  it equals to
\begin{equation}
\label{eq1.12.2}U_{ind}(0)= -\frac{e^2\omega_s}{2 v}\frac{\epsilon_0-1}{(\epsilon_0+1)}e^{-2\omega_sR_0/v}\mathrm{Ei}\left(2\frac{\omega_sR_0}{v}\right)
\end{equation}

In both cases considered, approaching the cavity surface, the induced potential increases. However, the induced potential must be cut off for the distances $R_0-r\sim q_0^{-1}$, therefore, the maximum value of the induced potential in both limits can be estimated as \footnote{In the following estimates, we set $\dfrac{\epsilon_0-1}{\epsilon_0+1}\sim1$.}
\begin{equation}
\label{eq1.12.1.3}U^{min}_{ind}\simeq \frac{e^2\omega_s}{v}\ln{\left(\frac{q_0v}{\omega_s}-1\right)}\ln{\left(\frac{\omega_s}{q_0v}\right)}\,,
\end{equation}
in which the logarithms do not essentially contribute to the value of induced potential. That allows us in further estimates to omit the logarithmic factors, i.e. $U^{min}_{ind}\sim e^2\omega_s/v$.

Eq.(\ref{eq1.11}) is obtained in view of smallness of the transverse energy in comparison with the longitudinal one, $E_{\parallel} \gg E_{\bot}$, which also imposes definite restrictions on the particle longitudinal velocity $v$. Supposing the maximum of transverse energy $E_{\bot}$ to be equal to the maximum of induced potential, we can write the limitation for longitudinal energy $E_{\parallel}\gg \left(e^2\omega_s\sqrt{m}\right)^{2/3}$ as well as for normalised longitudinal speed of a particle $\beta\equiv v/c \gg \left(e^2\omega_s/(mc^3)\right)^{1/3}$.

For example, for a channeled electron moving inside a cavity characterised by surface oscillations of a plasma frequency $\omega_s\sim 10^{15}$~c$^{- 1} $, the reduced potential is valid under the condition $E_{\parallel}\gg 1$~eV. In turn, it corresponds to the velocity constraint $10^{-3}\ll\beta <1$, where the upper limit should be correlated with the applicability of  nonrelativistic description.

As follows from Eq.(\ref {eq1.12.1.3}), the maximum value of induced potential depends mainly on the particle charge and its longitudinal velocity, resulting for electron in the estimate $U_{ind}^{max}\sim \beta^{-1}$ meV. 
Hence, the magnitude of the induced potential for one particle is not large in comparison with the averaged atomic potential $\bar{V}_{max}\sim10$~eV. However,  it might be essentially different for a beam (multi-particle) passing through the cavity.

\section{Conclusion}
In this paper, the problem of interaction of a charged particle with the inner surface of a cylindrical cavity in an infinite insulator has been analysed within definite analytical approach. Based on the Hamiltonian formalism, we have succeeded in reducing the interaction potential for a charged particle in the field of surface excitations. 

Neglecting excitations of the wavelengths comparable to the cavity radius, the interaction potential has been explicitly written revealing its complex nature. An imaginary part of the potential leads to a finite width of the energy levels and is not examined in general form. However, as an example, the energy losses of a particle per unit of traveled distance is obtained at its interaction with surface plasmons. The analysis of a real part of the potential, instead, has been carried out for two limiting cases.

We have shown for the first time that at the limit $\omega_sR_0/v\ll1$ the potential of interaction of a charged particle with the cavity surface acts as a scattering potential (forming a reflecting barrier), while at $\omega_sR_0/v\gg1$ it reveals a potential well near the surface.
The width of the potential well depends on the speed of the particle, i.e. the higher the speed of the particle the wider the well. In both cases considered, the real potential logarithmically tends to plus infinity. The maximum value of the induced potential mainly depends  on the particle charge and its longitudinal velocity. The estimates performed show that the averaged atomic potential is much higher than that induced for one particle, while for a beam of many particles channeled in a capillary the maximum value of the induced potential is expected to be essentially different.

\appendix
\section{The $g_q$-amplitudes}

To determine the amplitudes $g_q$ we solve the electrostatic task for a scalar potential of a charge $e$ inside an infinite cylindrical cavity
\begin{equation}
\label{eq2.1}\nabla^2\varphi=\left\{\begin{array}{ll}-4\pi e\delta(\mathbf{r}-\mathbf{r}_0),0\le r\le R_0 \\ 0,R_0<r\end{array}\right.
\end{equation}
valid at the following interface matching condition 
\begin{equation*}
\label{eq2.2}\varphi_{R_0+0}=\varphi_{R_0-0},\quad \left.\frac{\partial\varphi}{\partial r}\right|_{R_0-0}=\epsilon_0\left.\frac{\partial\varphi}{\partial r}\right|_{R_0+0},
\end{equation*}
where $\epsilon_0$ is the static dielectric constant. Presenting the potential  as a sum 
$\varphi=\sum\limits_{q,n}\tilde{\varphi}_n(q,r)e^{iqz+in\phi}$ we get at $\mathbf{r}_0=(0,r_0,0)$ the expression
\begin{equation}
\label{eq2.4}\tilde{\varphi}_n(q,r)=\left\{\begin{array}{lll}A_1K_{n}{qr},r\ge R_0\\ A_2K_n(qr)+B_2I_n(qr),R_0\ge r\ge r_0\\ A_3I_n(qr),r_0\ge r\ge 0\end{array}\right.
\end{equation}
with the following constants
\[A_1=\frac{e}{ L}I_n(qr_0)\left(1+\phantom{\frac{K_n}{K_n}}\right.\]
\begin{equation*}
\label{eq2.8}+\left.\frac{(\epsilon_0-1)K'_n(qR_0)I_n(qR_0)}{I_n'(qR_0)K_n(qR_0)-\epsilon_0I_n(qR_0)K'_n(qR_0)}\right),
\end{equation*}
\begin{equation*}
\label{eq2.5}A_2=\frac{e}{L}I_n(qr_0),
\end{equation*}
\[A_3=\frac{e}{ L}\left[K_n(qr_0)+\phantom{\frac{K_n}{K_n}}\right.\]
\begin{equation*}
\label{eq2.6}+\left.\frac{(\epsilon_0-1)K'_n(qR_0)K_n(qR_0)I_n(qr_0)}{I_n'(qR_0)K_n(qR_0)-\epsilon_0I_n(qR_0)K'_n(qR_0)}\right],
\end{equation*}
\begin{equation*}
\label{eq2.7}B_2=\frac{e}{L}\frac{(\epsilon_0-1)K'_n(qR_0)K_n(qR_0)I_n(qr_0)}{I_n'(qR_0)K_n(qR_0)-\epsilon_0I_n(qR_0)K'_n(qR_0)}
\end{equation*}

Correspondingly, the interaction energy of the induced charges with the particle can be presented as follows
\[U_{ind}=\frac{e^2}{ L}\times\]
\begin{equation}
\label{eq2.9}\times \sum\limits_{q,n}\frac{(\epsilon_0-1)K'_n(qR_0)K_n(qR_0)I^2_n(qr_0)}{I_n'(qR_0)K_n(qR_0)-\epsilon_0I_n(qR_0)K'_n(qR_0)}
\end{equation}
Comparing the this expression with Eq.(\ref{eq1.8.1}) at $k=0$, we can get the following definition for the amplitudes $g_q$  
\[\left|g_{qn}\right|^2=\left|g_{n}(qR_0)\right|^2=-\frac{e^2\omega_q}{ \hbar  L}\times\]
\begin{equation}
\label{eq2.10} \times \frac{(\epsilon_0-1)K'_n(qR_0)K_n(qR_0)}{I_n'(qR_0)K_n(qR_0)-\epsilon_0I_n(qR_0)K'_n(qR_0)}\,,
\end{equation}
which in a particular case of interactions with surface plasmons becomes equal to
\begin{equation}
\label{eq2.10}\left|g_{n}(qR_0)\right|^2_{\epsilon_0\to\infty}=\frac{e^2\omega_q}{ \hbar  L}\frac{K_n(qR_0)}{I_n(qR_0)}
\end{equation}
\label{app1}

\section{The sum calculation}

Analysing $e^{i(k_zz+n\phi)}$ for a plane wave propagating along the $zOy$-plane at $R_0\to\infty$, we can define $\phi\approx y/R_0$  near curved surface of very large radius that corresponds to the wave at  $n=k_yR_0$. Thus, we can rewrite the sum in Eq.(\ref{eq1.8.2}) in the following (below for brevity we omit the index  $y$ of the wave vector $k_y$)
\[\lim_{R_0\to\infty}\sum\limits_{n=-\infty}^{\infty}\frac{K_n\left(\frac{\omega_s}{v}R_0\right)}{I_n\left(\frac{\omega_s}{v}R_0\right)}I_n^2\left(\frac{\omega_s}{v}(R_0-x)\right)\approx\]
\begin{equation}
\label{eq3.0}\approx \lim_{R_0\to\infty}2\int\limits_0^{\infty}\frac{K_{kR_0}\left(z_1kR_0\right)}{I_{kR_0}\left(z_1kR_0\right)}I_{kR_0}^2\left(z_2kR_0\right)R_0dk
\end{equation}
\begin{equation*}
\label{eq3.1}z_1=\frac{\omega_s}{vk}; \quad z_2=z_1\left(1-\frac{x}{R_0}\right)
\end{equation*}
Neglecting small  $k\ll R_0^{-1}$ we present the Bessel functions in the form of asymptotic expansion \cite{abram} ($x/R_0\ll 1$)
\begin{equation*}
\label{eq3.2} K_{kR_0}\left(z_1kR_0\right)\approx \sqrt{\frac{\pi}{2kR_0}}\frac{e^{-kR_0\eta_1}}{(1+z_1^2)^{1/4}}
\end{equation*}
\begin{equation*}
\label{eq3.3} I_{kR_0}\left(z_1kR_0\right)\approx \frac{1}{\sqrt{2\pi kR_0}}\frac{e^{kR_0\eta_1}}{(1+z_1^2)^{1/4}}
\end{equation*}
\begin{equation*}
\label{eq3.4} I_{kR_0}\left(z_2kR_0\right)\approx \frac{1}{\sqrt{2\pi kR_0}}\frac{e^{kR_0\eta_2}}{(1+z_2^2)^{1/4}}
\end{equation*}
\begin{equation*}
\label{eq3.5} \eta_1=\sqrt{1+z_1^2}+\ln{\frac{z_1}{1+\sqrt{1+z_1^2}}}
\end{equation*}
\begin{equation*}
\label{eq3.6} \eta_2\approx \sqrt{1+z_1^2}+\ln{\frac{z_1}{1+\sqrt{1+z_1^2}}}-\frac{x}{R_0}\sqrt{1+z_1^2}
\end{equation*}
Substituting these expressions into Eq.(\ref{eq3.0}), we obtain
\[\lim_{R_0\to\infty}\sum\limits_{n=-\infty}^{\infty}\frac{K_n\left(\frac{\omega_s}{v}R_0\right)}{I_n\left(\frac{\omega_s}{v}R_0\right)}I_n^2\left(\frac{\omega_s}{v}(R_0-x)\right) =\]
\begin{equation}
\label{eq3.2}= \int\limits_0^{\infty}\frac{e^{-2x\sqrt{k^2+\omega_s^2/v^2}}}{\sqrt{k^2+\omega_s^2/v^2}}dk=K_0\left(2\frac{\omega_s}{v}x\right)
\end{equation}
Having assumed an infinite upper integration limit, while a finite - equal to $\sqrt{q_0^2-\omega_s^2/v^2}$, the result is conformed with that presented in \cite{jap}.
\label{app2}

\section{The sum and integral calculations}

In new notations $x=r/R_0$, $a=q_0R_0$ and $z_0=\omega_sR_0/v$ the induced potential is written in the form
\begin{equation}
\label{eq4.1}U_{ind}(r)=-\frac{e^2\omega_s}{2\pi v}F(x,a,z_0)\,,
\end{equation}
where 
\[F(x,a,z_0)=\ln{\left(\frac{a}{z_0}-1\right)}g(x,z_0)+\]
\begin{equation}
\label{eq4.2}+\sum\limits_{k=1}^{\infty}\frac{(a-z_0)^k-(-z_0)^k}{kk!}\frac{\partial^k}{\partial z_0^k}g(x,z_0)
\end{equation}
\begin{equation}
\label{eq4.3}g(x,z_0)=\sum\limits_{n=-\infty}^{\infty}\frac{(\epsilon_0-1)K_n'(z_0)K_n(z_0)I_n^2(xz_0)}{I_n'(z_0)K_n(z_0)-\epsilon_0 K_n'(z_0)I_n(z_0)}
\end{equation}

\subsection{$z_0\gg 1$:}
For this limit, we can use the asymptotic expressions for the Bessel functions \cite{abram}. Then at $0\ll x\le 1$ we can calculate the sum by replacing it with the integral, which successfully can be simplified within the saddle-point method

\[g(x,z_0)\approx -\frac{\epsilon_0-1}{2(\epsilon_0+1)}\times\]
\begin{equation}
\label{eq4.4}\times e^{-2(1-x)(1-x/4)z_0}K_0\left(\frac{(1-x)x}{2}z_0\right)
\end{equation}
Substituting (\ref{eq4.4}) into (\ref{eq4.2}), we obtain

\[F(x,a,z_0)= -\frac{\epsilon_0-1}{2(\epsilon_0+1)}e^{-z_0\alpha(x)}\times\]
\[\times\int\limits_{0}^{\infty}\left(\mathrm{Ei}\left[(z_0-a)(\alpha(x)+\beta(x)\mathrm{ch}{(t)})\right]\right.-\]
\begin{equation}
\label{eq4.5}-\left.\mathrm{Ei}\left[z_0(\alpha(x)+\beta(x)\mathrm{ch}{(t)})\right]\right)e^{-z_0\beta(x)\mathrm{ch}{(t)}}dt \,,
\end{equation}
where $\alpha(x)=2(1-x)(1-x/4),\beta(x)=(1-x)x/2$. For $x\to 1$, $(\alpha(x),\beta(x))\to 0$, that corresponds to the postion at the surface, we get the function $F(x,a,z_0)$  as follows
\[F(x,a,z_0)\approx-\frac{\epsilon_0-1}{2(\epsilon_0+1)}e^{-z_0\alpha(x)}\times\]
\begin{equation}
\label{eq4.6}\times\ln{\left(\frac{a}{z_0}-1\right)}K_0\left(\beta(x)z_0\right) 
\end{equation}
This result proves, that approaching the surface, the potential tends logarithmically to minus infinity. If $a\gg z_0$, then the 1st term in (\ref{eq4.5}) can be neglected, and the integral itself very weakly depends on the functions $\alpha(x)$ and $\beta(x)$. Thus, rather precisely the expression (\ref{eq4.5}) can be replaced with
\begin{equation}
\label{eq4.7} F(x,a,z_0)= \frac{\epsilon_0-1}{2(\epsilon_0+1)}G\left(\frac{3}{2}z_0(1-x)\right)\,,
\end{equation}
where
\begin{equation*}
\label{eq4.8} G(\xi)=\int\limits_{0}^{\infty}\mathrm{Ei}\left(\xi\left(1+\frac{\mathrm{ch}{(t)}}{3}\right)\right)e^{-\xi(1+\mathrm{ch}{(t)}/3)}dt
\end{equation*}
The reduced function has a maximum $G(\xi_m)=2.59815$ at $\xi_m=0.545901$. It is notable that in the limiting case $R_0\to\infty$ the form of the potential remains unchanged and is determined by the function $G(3\omega_sx/2v)$ ($x$ is the distance from the particle to the plane).

\subsection{$z_0\ll 1$:}

At $z_0\ll1$, using expressions for Bessel functions of a small argument, we obtain the following expression for the function $g (x,z_0)$

\begin{equation}
\label{eq4.9} g(x,z_0)\approx \frac{\epsilon_0-1}{\epsilon_0+1}\ln{\left(1-x^2\right)}+\frac{\epsilon_0-1}{\epsilon_0}\left[\ln{\left(\frac{z_0}{2}\right)}+\gamma_0\right]\,,
\end{equation}
where $\gamma_0$ is the Euler's constant. Substitution of (\ref{eq4.9}) into (\ref{eq4.2}) followed by summing results in
\[F(x,a,z_0)=\frac{\epsilon_0-1}{\epsilon_0+1}\ln{\left(\frac{a}{z_0}-1\right)}\ln{\left(1-x^2\right)}\]
\begin{equation}
\label{eq4.10}   + C(\epsilon_0,a,z_0)
\end{equation}
with
\[C(\epsilon_0,a,z_0)=\frac{\epsilon_0-1}{\epsilon_0}\left[\ln{\left(\frac{a}{z_0}-1\right)}\left(\ln{\left(\frac{z_0}{2}\right)}+\gamma_0\right)\right.\]
\begin{equation}
\label{eq4.11} \left.-\left(\mathrm{Li}_2\left(1-\frac{a}{z_0}\right)-\mathrm{Li}_2\left(1\right)\right)\right]
\end{equation}
Here $\mathrm{Li}_2(\xi)$ is the polylogarithm function. As seen from (\ref{eq4.10}), the interaction potential at $z_0\ll1$ is attractive with the tendency of logarithmic dropp-off when the projectile approaches the cavity wall.
\label{app3}

\bibliography{mybibfile}

\begin{thebibliography}{22}%
\makeatletter
\providecommand \@ifxundefined [1]{%
 \@ifx{#1\undefined}
}%
\providecommand \@ifnum [1]{%
 \ifnum #1\expandafter \@firstoftwo
 \else \expandafter \@secondoftwo
 \fi
}%
\providecommand \@ifx [1]{%
 \ifx #1\expandafter \@firstoftwo
 \else \expandafter \@secondoftwo
 \fi
}%
\providecommand \natexlab [1]{#1}%
\providecommand \enquote  [1]{``#1''}%
\providecommand \bibnamefont  [1]{#1}%
\providecommand \bibfnamefont [1]{#1}%
\providecommand \citenamefont [1]{#1}%
\providecommand \href@noop [0]{\@secondoftwo}%
\providecommand \href [0]{\begingroup \@sanitize@url \@href}%
\providecommand \@href[1]{\@@startlink{#1}\@@href}%
\providecommand \@@href[1]{\endgroup#1\@@endlink}%
\providecommand \@sanitize@url [0]{\catcode `\\12\catcode `\$12\catcode
  `\&12\catcode `\#12\catcode `\^12\catcode `\_12\catcode `\%12\relax}%
\providecommand \@@startlink[1]{}%
\providecommand \@@endlink[0]{}%
\providecommand \url  [0]{\begingroup\@sanitize@url \@url }%
\providecommand \@url [1]{\endgroup\@href {#1}{\urlprefix }}%
\providecommand \urlprefix  [0]{URL }%
\providecommand \Eprint [0]{\href }%
\providecommand \doibase [0]{https://doi.org/}%
\providecommand \selectlanguage [0]{\@gobble}%
\providecommand \bibinfo  [0]{\@secondoftwo}%
\providecommand \bibfield  [0]{\@secondoftwo}%
\providecommand \translation [1]{[#1]}%
\providecommand \BibitemOpen [0]{}%
\providecommand \bibitemStop [0]{}%
\providecommand \bibitemNoStop [0]{.\EOS\space}%
\providecommand \EOS [0]{\spacefactor3000\relax}%
\providecommand \BibitemShut  [1]{\csname bibitem#1\endcsname}%
\let\auto@bib@innerbib\@empty
\bibitem [{\citenamefont {Shiltsev}\ and\ \citenamefont
  {Zimmermann}(2021)}]{shizim-rmp2021}%
  \BibitemOpen
  \bibfield  {author} {\bibinfo {author} {\bibfnamefont {V.}~\bibnamefont
  {Shiltsev}}\ and\ \bibinfo {author} {\bibfnamefont {F.}~\bibnamefont
  {Zimmermann}},\ }\href {https://doi.org/10.1103/RevModPhys.93.015006}
  {\bibfield  {journal} {\bibinfo  {journal} {Rev. Mod. Phys.}\ }\textbf
  {\bibinfo {volume} {93}},\ \bibinfo {pages} {015006} (\bibinfo {year}
  {2021})}\BibitemShut {NoStop}%
\bibitem [{\citenamefont {Minty}\ and\ \citenamefont {Zimmermann}(5
  21)}]{mizi-book2003}%
  \BibitemOpen
  \bibfield  {author} {\bibinfo {author} {\bibfnamefont {M.~G.}\ \bibnamefont
  {Minty}}\ and\ \bibinfo {author} {\bibfnamefont {F.}~\bibnamefont
  {Zimmermann}},\ }\href {https://doi.org/10.1007/978-3-662-08581-3{\_}13}
  {\emph {\bibinfo {title} {Measurement and Control of Charged Particle
  Beams}}},\ Particle Acceleration and Detection\ (\bibinfo  {publisher}
  {Springer},\ \bibinfo {address} {Berlin, Heidelberg},\ \bibinfo {year}
  {2003-05-21})\BibitemShut {NoStop}%
\bibitem [{\citenamefont {Zimmermann}\ \emph {et~al.}(2021)\citenamefont
  {Zimmermann}, \citenamefont {Seidling},\ and\ \citenamefont
  {Hommelhoff}}]{ziseho-nature2021}%
  \BibitemOpen
  \bibfield  {author} {\bibinfo {author} {\bibfnamefont {R.}~\bibnamefont
  {Zimmermann}}, \bibinfo {author} {\bibfnamefont {M.}~\bibnamefont
  {Seidling}},\ and\ \bibinfo {author} {\bibfnamefont {P.}~\bibnamefont
  {Hommelhoff}},\ }\href {https://doi.org/10.1038/s41467-020-20592-4}
  {\bibfield  {journal} {\bibinfo  {journal} {Nature Communications}\ }\textbf
  {\bibinfo {volume} {12}},\ \bibinfo {pages} {390} (\bibinfo {year}
  {2021})}\BibitemShut {NoStop}%
\bibitem [{\citenamefont {Gras-Marti}\ \emph {et~al.}(1991)\citenamefont
  {Gras-Marti}, \citenamefont {Urbassek}, \citenamefont {Arista},\ and\
  \citenamefont {Flores}}]{nato-asi1991}%
  \BibitemOpen
  \bibinfo {editor} {\bibfnamefont {A.}~\bibnamefont {Gras-Marti}}, \bibinfo
  {editor} {\bibfnamefont {H.~M.}\ \bibnamefont {Urbassek}}, \bibinfo {editor}
  {\bibfnamefont {N.~R.}\ \bibnamefont {Arista}},\ and\ \bibinfo {editor}
  {\bibfnamefont {F.}~\bibnamefont {Flores}},\ eds.,\ \href
  {https://doi.org/https://doi.org/10.1007/978-1-4684-8026-9} {\emph {\bibinfo
  {title} {Interaction of Charged Particles with Solids and Surfaces}}},\
  \bibinfo {series} {Nato ASI Series}, Vol.\ \bibinfo {volume} {271}\ (\bibinfo
   {publisher} {Springer, Boston, MA},\ \bibinfo {year} {1991})\BibitemShut
  {NoStop}%
\bibitem [{\citenamefont {Dabagov}(2003)}]{dabagov-physusp2003}%
  \BibitemOpen
  \bibfield  {author} {\bibinfo {author} {\bibfnamefont {S.~B.}\ \bibnamefont
  {Dabagov}},\ }\href {https://doi.org/10.1070/pu2003v046n10abeh001639}
  {\bibfield  {journal} {\bibinfo  {journal} {Physics-Uspekhi}\ }\textbf
  {\bibinfo {volume} {46}},\ \bibinfo {pages} {1053} (\bibinfo {year}
  {2003})}\BibitemShut {NoStop}%
\bibitem [{\citenamefont {Kumakhov}(1986)}]{kumakhov-bookinru1986}%
  \BibitemOpen
  \bibfield  {author} {\bibinfo {author} {\bibfnamefont {M.~A.}\ \bibnamefont
  {Kumakhov}},\ }\href@noop {} {\emph {\bibinfo {title} {Radiation of Channeled
  Particles in Crystals}}},\ \bibinfo {edition} {{Russian}}\ ed.\ (\bibinfo
  {publisher} {Energoatomizdat},\ \bibinfo {year} {1986})\BibitemShut {NoStop}%
\bibitem [{\citenamefont {Kumakhov}\ and\ \citenamefont
  {Komarov}(1990)}]{kuko-physrep1990}%
  \BibitemOpen
  \bibfield  {author} {\bibinfo {author} {\bibfnamefont {M.}~\bibnamefont
  {Kumakhov}}\ and\ \bibinfo {author} {\bibfnamefont {F.}~\bibnamefont
  {Komarov}},\ }\href {https://doi.org/10.1016/0370-1573(90)90135-O} {\bibfield
   {journal} {\bibinfo  {journal} {Physics Reports}\ }\textbf {\bibinfo
  {volume} {191}},\ \bibinfo {pages} {289} (\bibinfo {year}
  {1990})}\BibitemShut {NoStop}%
\bibitem [{\citenamefont {Khounsary}\ and\ \citenamefont
  {MacDonald}(2010)}]{khomac-xoi2010}%
  \BibitemOpen
  \bibfield  {author} {\bibinfo {author} {\bibfnamefont {A.}~\bibnamefont
  {Khounsary}}\ and\ \bibinfo {author} {\bibfnamefont {C.~A.}\ \bibnamefont
  {MacDonald}},\ }\href {https://doi.org/10.1155/2010/867049} {\bibfield
  {journal} {\bibinfo  {journal} {X-Ray Optics and Instrumentation}\ }\textbf
  {\bibinfo {volume} {2010}},\ \bibinfo {pages} {867049} (\bibinfo {year}
  {2010})}\BibitemShut {NoStop}%
\bibitem [{\citenamefont {Dabagov}\ and\ \citenamefont
  {Gladkikh}(2019)}]{dagl-rpc2019}%
  \BibitemOpen
  \bibfield  {author} {\bibinfo {author} {\bibfnamefont {S.~B.}\ \bibnamefont
  {Dabagov}}\ and\ \bibinfo {author} {\bibfnamefont {Y.~P.}\ \bibnamefont
  {Gladkikh}},\ }\href
  {https://doi.org/https://doi.org/10.1016/j.radphyschem.2018.06.009}
  {\bibfield  {journal} {\bibinfo  {journal} {Radiation Physics and Chemistry}\
  }\textbf {\bibinfo {volume} {154}},\ \bibinfo {pages} {3 } (\bibinfo {year}
  {2019})}\BibitemShut {NoStop}%
\bibitem [{\citenamefont {Stolterfoht}\ \emph {et~al.}(2002)\citenamefont
  {Stolterfoht}, \citenamefont {Bremer}, \citenamefont {Hoffmann},
  \citenamefont {Hellhammer}, \citenamefont {Fink}, \citenamefont {Petrov},\
  and\ \citenamefont {Sulik}}]{stol}%
  \BibitemOpen
  \bibfield  {author} {\bibinfo {author} {\bibfnamefont {N.}~\bibnamefont
  {Stolterfoht}}, \bibinfo {author} {\bibfnamefont {J.-H.}\ \bibnamefont
  {Bremer}}, \bibinfo {author} {\bibfnamefont {V.}~\bibnamefont {Hoffmann}},
  \bibinfo {author} {\bibfnamefont {R.}~\bibnamefont {Hellhammer}}, \bibinfo
  {author} {\bibfnamefont {D.}~\bibnamefont {Fink}}, \bibinfo {author}
  {\bibfnamefont {A.}~\bibnamefont {Petrov}},\ and\ \bibinfo {author}
  {\bibfnamefont {B.}~\bibnamefont {Sulik}},\ }\href
  {https://doi.org/10.1103/PhysRevLett.88.133201} {\bibfield  {journal}
  {\bibinfo  {journal} {Phys. Rev. Lett.}\ }\textbf {\bibinfo {volume} {88}},\
  \bibinfo {pages} {133201} (\bibinfo {year} {2002})}\BibitemShut {NoStop}%
\bibitem [{\citenamefont {Schiessl}\ \emph
  {et~al.}(2005{\natexlab{a}})\citenamefont {Schiessl}, \citenamefont
  {Palfinger}, \citenamefont {T\ifmmode~\mbox{\H{o}}\else \H{o}\fi{}k\'esi},
  \citenamefont {Nowotny}, \citenamefont {Lemell},\ and\ \citenamefont
  {Burgd}}]{schi}%
  \BibitemOpen
  \bibfield  {author} {\bibinfo {author} {\bibfnamefont {K.}~\bibnamefont
  {Schiessl}}, \bibinfo {author} {\bibfnamefont {W.}~\bibnamefont {Palfinger}},
  \bibinfo {author} {\bibfnamefont {K.}~\bibnamefont
  {T\ifmmode~\mbox{\H{o}}\else \H{o}\fi{}k\'esi}}, \bibinfo {author}
  {\bibfnamefont {H.}~\bibnamefont {Nowotny}}, \bibinfo {author} {\bibfnamefont
  {C.}~\bibnamefont {Lemell}},\ and\ \bibinfo {author} {\bibfnamefont
  {J.}~\bibnamefont {Burgd}},\ }\href
  {https://doi.org/10.1103/PhysRevA.72.062902} {\bibfield  {journal} {\bibinfo
  {journal} {Phys. Rev. A}\ }\textbf {\bibinfo {volume} {72}},\ \bibinfo
  {pages} {062902} (\bibinfo {year} {2005}{\natexlab{a}})}\BibitemShut
  {NoStop}%
\bibitem [{\citenamefont {Stolterfoht}\ and\ \citenamefont
  {Yamazaki}(2016)}]{sto-ya_phrep2016}%
  \BibitemOpen
  \bibfield  {author} {\bibinfo {author} {\bibfnamefont {N.}~\bibnamefont
  {Stolterfoht}}\ and\ \bibinfo {author} {\bibfnamefont {Y.}~\bibnamefont
  {Yamazaki}},\ }\href
  {https://doi.org/https://doi.org/10.1016/j.physrep.2016.02.008} {\bibfield
  {journal} {\bibinfo  {journal} {Physics Reports}\ }\textbf {\bibinfo {volume}
  {629}},\ \bibinfo {pages} {1} (\bibinfo {year} {2016})}\BibitemShut {NoStop}%
\bibitem [{\citenamefont {Schiessl}\ \emph
  {et~al.}(2005{\natexlab{b}})\citenamefont {Schiessl}, \citenamefont
  {Palfinger}, \citenamefont {Lemell},\ and\ \citenamefont
  {Burgd{\"o}rfer}}]{schi1}%
  \BibitemOpen
  \bibfield  {author} {\bibinfo {author} {\bibfnamefont {K.}~\bibnamefont
  {Schiessl}}, \bibinfo {author} {\bibfnamefont {W.}~\bibnamefont {Palfinger}},
  \bibinfo {author} {\bibfnamefont {C.}~\bibnamefont {Lemell}},\ and\ \bibinfo
  {author} {\bibfnamefont {J.}~\bibnamefont {Burgd{\"o}rfer}},\ }\href
  {https://doi.org/https://doi.org/10.1016/j.nimb.2005.03.050} {\bibfield
  {journal} {\bibinfo  {journal} {Nucl. Instr. Meth. B}\ }\textbf {\bibinfo
  {volume} {232}},\ \bibinfo {pages} {228} (\bibinfo {year}
  {2005}{\natexlab{b}})},\ \bibinfo {note} {inelastic Ion-Surface
  Collisions}\BibitemShut {NoStop}%
\bibitem [{\citenamefont {Mahan}(1974)}]{amer}%
  \BibitemOpen
  \bibfield  {author} {\bibinfo {author} {\bibfnamefont {G.~D.}\ \bibnamefont
  {Mahan}},\ }\bibinfo {title} {Elementary excitations in solids, molecules and
  atoms. part b}\ (\bibinfo  {publisher} {Springer US},\ \bibinfo {year}
  {1974})\ Chap.\ \bibinfo {chapter} {Electron Interaction with Surface
  Modes}\BibitemShut {NoStop}%
\bibitem [{\citenamefont {{O. Madelung}}(1980)}]{madelung}%
  \BibitemOpen
  \bibfield  {author} {\bibinfo {author} {\bibnamefont {{O. Madelung}}},\
  }\href@noop {} {\emph {\bibinfo {title} {Solid State Theory}}},\ \bibinfo
  {edition} {{Russian}}\ ed.,\ ( Translated from O. Madelung,
  Festk\"{or}pertheorie I, II, Spriger-Verlag 1972)\ (\bibinfo  {publisher}
  {Nauka},\ \bibinfo {year} {1980})\BibitemShut {NoStop}%
\bibitem [{\citenamefont {Ashkroft}\ and\ \citenamefont {Mermin}(1980)}]{ash}%
  \BibitemOpen
  \bibfield  {author} {\bibinfo {author} {\bibfnamefont {N.}~\bibnamefont
  {Ashkroft}}\ and\ \bibinfo {author} {\bibfnamefont {N.}~\bibnamefont
  {Mermin}},\ }\href@noop {} {\emph {\bibinfo {title} {Solid State Physics}}},\
  \bibinfo {edition} {{Russian}}\ ed.,\ (Translated from N.W. Ashkroft, N.D.
  Mermin, Solid State Physics, Saunders College Publ. 1976)\ (\bibinfo
  {publisher} {Nauka},\ \bibinfo {year} {1980})\BibitemShut {NoStop}%
\bibitem [{\citenamefont {Dedkov}(1995)}]{ded}%
  \BibitemOpen
  \bibfield  {author} {\bibinfo {author} {\bibfnamefont {G.~V.}\ \bibnamefont
  {Dedkov}},\ }\href {https://doi.org/10.1070/PU1995v038n08ABEH000100}
  {\bibfield  {journal} {\bibinfo  {journal} {Physics-Uspekhi}\ }\textbf
  {\bibinfo {volume} {38}},\ \bibinfo {pages} {877} (\bibinfo {year}
  {1995})}\BibitemShut {NoStop}%
\bibitem [{\citenamefont {Kawai}\ \emph {et~al.}(1982)\citenamefont {Kawai},
  \citenamefont {Itoh},\ and\ \citenamefont {Ohtsuki}}]{jap}%
  \BibitemOpen
  \bibfield  {author} {\bibinfo {author} {\bibfnamefont {R.}~\bibnamefont
  {Kawai}}, \bibinfo {author} {\bibfnamefont {N.}~\bibnamefont {Itoh}},\ and\
  \bibinfo {author} {\bibfnamefont {Y.}~\bibnamefont {Ohtsuki}},\ }\href
  {https://doi.org/https://doi.org/10.1016/0039-6028(82)90461-7} {\bibfield
  {journal} {\bibinfo  {journal} {Surface Science}\ }\textbf {\bibinfo {volume}
  {114}},\ \bibinfo {pages} {137} (\bibinfo {year} {1982})}\BibitemShut
  {NoStop}%
\bibitem [{\citenamefont {Nunez}\ \emph {et~al.}(1980)\citenamefont {Nunez},
  \citenamefont {Echenique},\ and\ \citenamefont {Ritchie}}]{amer2}%
  \BibitemOpen
  \bibfield  {author} {\bibinfo {author} {\bibfnamefont {R.}~\bibnamefont
  {Nunez}}, \bibinfo {author} {\bibfnamefont {P.}~\bibnamefont {Echenique}},\
  and\ \bibinfo {author} {\bibfnamefont {R.}~\bibnamefont {Ritchie}},\ }\href
  {http://iopscience.iop.org/0022-3719/13/22/017} {\bibfield  {journal}
  {\bibinfo  {journal} {J. Phys. C: Solid State Phys.}\ }\textbf {\bibinfo
  {volume} {13}},\ \bibinfo {pages} {4229} (\bibinfo {year}
  {1980})}\BibitemShut {NoStop}%
\bibitem [{\citenamefont {Lindhard}(1965)}]{lindhard1965}%
  \BibitemOpen
  \bibfield  {author} {\bibinfo {author} {\bibfnamefont {J.}~\bibnamefont
  {Lindhard}},\ }\href {https://www.osti.gov/biblio/4536390} {\bibfield
  {journal} {\bibinfo  {journal} {Kgl. Dan. Vid. Selsk. Mat.- Fys. Medd.}\
  }\textbf {\bibinfo {volume} {34}} (\bibinfo {year} {1965})}\BibitemShut
  {NoStop}%
\bibitem [{\citenamefont {Gemmell}(1974)}]{gemmell-rmp1974}%
  \BibitemOpen
  \bibfield  {author} {\bibinfo {author} {\bibfnamefont {D.~S.}\ \bibnamefont
  {Gemmell}},\ }\href {https://doi.org/10.1103/RevModPhys.46.129} {\bibfield
  {journal} {\bibinfo  {journal} {Rev. Mod. Phys.}\ }\textbf {\bibinfo {volume}
  {46}},\ \bibinfo {pages} {129} (\bibinfo {year} {1974})}\BibitemShut
  {NoStop}%
\bibitem [{\citenamefont {Abramowitz}\ and\ \citenamefont
  {I.A.~Stegun}(1979)}]{abram}%
  \BibitemOpen
  \bibfield  {author} {\bibinfo {author} {\bibfnamefont {M.}~\bibnamefont
  {Abramowitz}}\ and\ \bibinfo {author} {\bibfnamefont {I.}~\bibnamefont
  {I.A.~Stegun}},\ }\href@noop {} {\emph {\bibinfo {title} {Handbook of
  Mathematical Functions}}},\ (M. Abramowitz, I.A. Stegun, Handbook of
  Mathematical Functions, Dover, NY 1971)\ (\bibinfo  {publisher} {Nauka},\
  \bibinfo {year} {1979})\BibitemShut {NoStop}%
\end{thebibliography}%

\end{document}